\documentclass[aps,prl,preprint,groupedaddress]{revtex4}
\usepackage{graphicx}

\begin{document}


\title{Electrical transport across Au/Nb:STO Schottky interface with different Nb doping}
\author{K. G. Rana, V. Khikhlovskyi and T. Banerjee}
 \email{T.Banerjee@rug.nl}
 \affiliation{
Physics of Nanodevices, Zernike Institute for Advanced Materials, University of Groningen, Groningen, 
The Netherlands
}%

\date{\today}

\begin{abstract} 

We have investigated electron transport in Nb doped SrTiO$_3$ single crystals for two doping densities.
We find that the resistivity and mobility are temperature dependent in both whereas the carrier concentration is almost temperature invariant. We  rationalize this using the hydrogenic theory for shallow donors. Further, we probe 
electrical transport across Schottky interfaces of Au on TiO$_2$ terminated n-type SrTiO$_3$. Quantitative analysis of macroscopic I-V 
measurements reveal thermionic emission dominated transport for the low doped substrate whereas it deviates from such behavior for the high doped substrate. This work is relevant for designing devices to study electronic transport using oxide-semiconductors. 

\end{abstract}


\maketitle

Probing electron transport across a Schottky interface between a metal and an oxide semiconductor is vital to the development of oxide electronics. Such studies can yield important information on the mechanisms prevalent in interface transport and are relevant to devise methods and engineer energy band profiles in devices exploiting these materials. Electrical transport across conventional semiconductor(as Si)/metal Schottky interface is well established and described by thermionic emission theory for moderately doped semiconductors, whereas for heavily doped semiconductors, quantum mechanical tunneling is the dominant transport process.\cite{M.Sze} Although such physics of charge transport across a Schottky interface between metal and Si is well understood, the underlying mechanism of electrical transport across an interface with oxide semiconductor as doped SrTiO$_3$ is not. 

In this context, it should be mentioned that transport in oxide heterostructures with undoped SrTiO$_3$ (STO) has led to the observation of unparalleled physical phenomena in condensed matter.\cite{Ohtomo, Mannhart, Brinkman, Herranz} However, research related to transport in electronic devices utilizing doped STO has recently gained momentum. SrTiO$_3$ is a band gap insulator and it's electrical conductivity can be tuned by either substituting La$^{3+}$ (acceptor) for Sr$^{2+}$ or Nb$^{5+}$ (donor) for Ti$^{4+}$ giving rise to p and n type conduction respectively. \cite{S.Ohta}

In this work, we have studied electronic transport in doped STO for Nb doping of 0.01 wt $\%$ and 0.1 wt $\%$ (Nb:STO), over the temperature range 100 K to 300 K. For both cases, we find an almost temperature invariant carrier concentration but a strong temperature dependent resistivity exhibiting a power law dependence (unlike that found in conventional semiconductors). A hydrogenic theory of shallow donors with a temperature dependent dielectric permittivity in STO explains the behavior well. Further, using such well-characterized substrates, electrical transport across an interface with a high work function material as Au, was studied at room temperature (RT), using current-voltage ($I-V$) measurement. Analysis of the forward-bias characteristic reveals an exponential dependence, consistent with thermionic emission over the Schottky barrier and negligible current in the reverse direction, at RT. However, deviation from this behavior is observed for transport across the Schottky interface with 0.1 wt $\%$ Nb doping and accompanied by a larger reverse-bias current. The $I-V$ characteristics for both the Schottky interfaces exhibit a hysteretic feature. 

For this work, we used commercially available 5$\times$5 mm$^{2}$ Nb:STO (001) single crystals with 0.01 wt $\%$ and 0.1 wt $\%$ of Nb doping. In all cases, the substrates were treated using a standard chemical protocol \cite{G.KosterB} followed by annealing at 950 $^{\circ}$C for 1 h. Atomically flat singly terminated TiO$_2$ surface was confirmed from AFM studies. The resistivity and carrier concentration measurements were done using a four-probe van der Pauw geometry with Ti(200 nm)/Au(100 nm) contacts. For electrical transport measurements, Au (20 nm) was evaporated on TiO$_2$ terminated Nb:STO substrates and fabricated into Au/Nb:STO devices using standard UV-lithography and wet etching. The devices are of dimension 250$\times$650 $\mu$m$^{2}$. Ohmic contacts were realized by sputtering Ti/Au to the back of Nb:STO substrate.

The temperature dependence of resistivity and Hall coefficient were measured in the van der Pauw configuration, using a Physical Property Measuring System (PPMS by Quantum Design) from 100 K to 300 K. Figure 1 shows the variation of $\rho$ (resistivity) with temperature in Nb:STO for 0.01 and 0.1 wt $\%$ Nb doping. A few interesting observations emerge from this $\rho$-T dependence. First, $\rho$ decreases significantly with decreasing temperature strongly following a power law. Further, with increasing Nb doping, $\rho$ is observed to decrease while maintaining a positive $d\rho$/$dT$ upto 100 K. The 
monotonically decreasing $\rho$ with T is fitted with a generic power law as $\rho={A\times T^{-\alpha}}$, where, A is a temperature independent parameter and $\alpha$ is found to be 2.83 and 2.70 for 0.01 wt $\%$ and 0.1 wt $\%$ Nb doping respectively. Such a metal-like behavior of electrical resistivity is typical of an unconventional heavily doped semiconductors. The obtained resistivity values are in good agreement with earlier reports.\cite{A.Spinelli, H.P.R.Frederikse, O.N.Tufte}
In-plane electronic transport was further analyzed to determine the sign of the charge carrier and its concentration. All measurements were done in the linear current-voltage regime. A negative sign of the Hall coefficient (R$_H$=(1/q$N_D$)), consistent with electrons being the charge carriers ($N_D$), was obtained. 

Temperature dependence of the carrier concentration obtained for 0.01 and 0.1 wt $\%$ of Nb doping, in the temperature range 100 K to 300 K, is shown in Figure 2. We find an almost temperature invariant $N_D$ in both cases. We rationalize this using the hydrogenic theory of shallow donors where the donor binding energy, $E_D$, is related to the dielectric constant ($\epsilon$), the effective mass ($m_e^*$) and free-electron mass ($m_e$) as, $E_D = -13.6 eV(\frac{m_e^*}{m_e})(\frac{1}{\epsilon_r^2})$. 
Due to the large temperature dependence of $\epsilon$ in STO \cite{Barrett} (assumed to be the same in Nb:STO), the activation energy is small, even at 100 K. Thus, E$_D$ at 300 K is $\sim$ 1.5 meV decreasing to only $\sim$ 0.2 $\mu$eV at 4.2 K.\cite{A.Spinelli} Unlike what is observed in conventional semiconductors, this thus indicates, that the shallow donors remain ionized at all temperatures measured, implying the absence of carrier freeze out and explaining an almost temperature independence of the carrier concentration. However, we have observed a slight increase in the carrier concentration with increasing temperature, for both dopant densities (a factor of 1.4 in 0.01 wt $\%$ and 2 in 0.1 wt $\%$ of Nb doping). This could be ascribed to additional conduction processes originating from charge carriers at energy levels below the conduction band whose temperature dependence could be different from the carriers at the bottom of the conduction band in Nb:STO.\cite {C.LeePRB71}

The resistivity and carrier density thus obtained are used to extract the mobility of the charge carriers in Nb:STO using $\rho(T) = \frac{1}{\mu(T)q N_D(T)}$. Figure 2 (inset) shows the temperature dependence of mobility for both cases. The mobility is observed to decrease with increasing temperature. The temperature dependence of mobility is also fitted with a power law as $\mu = \mu_o\times T^{-m}$ where, $\mu_o$ is a temperature independent coefficient and the exponent m is related to the  different scattering mechanisms: electron-electron, electron-ionized impurity and electron-phonon scattering. Simple fitting of the obtained data gives m = 3.2 and 3 for Nb doping of 0.01 and of 0.1 wt $\%$ respectively. The temperature dependence of $\mu$(T) suggests the role of electron-phonon scattering due to the strong ionic nature of the lattice. It could be related to the temperature dependence of $\epsilon$, which implies effective screening of the ionized impurity scattering centers resulting in a metal-like behavior.\cite{H.P.R.Frederikse, O.N.Tufte}

Further, we have studied the electrical transport across an interface between a noble metal as Au and such well characterized Nb:STO substrates. Earlier studies on such substrates highlighted the necessity of special surface preparation protocols for studying electrical transport across such interface.\cite{T.Shimizu, T.Susaki, Y.Hikita} However, we use singly terminated (TiO$_2$) Nb:STO substrates, with a large area Ti/Au ohmic contact to study electrical transport across the interface between Au and Nb:STO at RT. Figure 3 (a) and (b) shows the current-voltage $(I-V)$ characteristics, at RT, of Au/Nb:STO Schottky diodes with 0.01 and of 0.1 wt $\%$ Nb doping. The plots are representative of measurements performed on simultaneously processed 10-12 diodes, for each doping concentration. Forward bias is defined when a positive bias is applied to Au. The $I-V$ characteristic in Fig. 3 (a) shows a clear rectifying behavior at RT with the sign of rectification as expected for a Schottky junction between a n-type semiconductor and metal. The forward bias characteristic on a semi-log scale shows a linear increase with bias whereas almost no current flows in the reverse bias. The linear increase and strong asymmetry in $I$ is consistent with thermionic emission over the Schottky barrier. Fig.3 (b) shows the $(I-V)$ characteristic for Au on 0.1 wt $\%$ Nb:STO at RT. The forward bias characteristic of the high doped semiconducting substrate also shows a steep increase in current with bias. However, a comparatively larger and bias-dependent reverse saturation current is seen here, unlike that observed in the low doped substrate. The hysteretic behavior in $(I-V)$ for both Au/Nb:STO diodes can be ascribed to the redistribution of oxygen vacancies or defects at the metal-semiconductor (M/S) interface with the sweeping of bias.\cite{SawaTokura}  

The Schottky barrier height (SBH) at zero bias and the ideality factor (n) was obtained from the $I-V$ plots by fitting the forward bias characteristics using the thermionic emission theory \cite{M.Sze} for both cases:
\begin{equation}
I=A^{*}AT^2exp(-\frac{q\phi_B}{k_BT})\left[exp(\frac{qV}{nk_BT})-1\right]
\end{equation}
Here $A^*$, the Richardson constant is assumed to be 156 Acm$^{-2}$K$^{-2}$, and $\phi_B$ is the zero bias SBH. All other symbols have their usual meaning \cite{M.Sze}. The zero bias SBH is found to be 1.15$\pm$0.05 eV (0.78$\pm$0.03 eV) with n = 1.1 (2) for the 0.01 (0.1) wt $\%$ of Nb doping in Au/Nb:STO diodes. The simple Schottky-Mott model for a M/S junction states that $\phi_B$, the SBH, is a difference of the bulk work function $W_o$ in the metal and $\chi$, the electron affinity in the semiconductor. Considering $W_o$ in bulk Au to be 5.1 eV and $\chi$ in Nb:STO to be 4 eV \cite{J.Robertson}, the SBH, $\phi_B$, is $\sim$ 1.1 eV. Such a close match of the SBH obtained from our experiments with that of this simple model, for the low doped Au/Nb:STO Schottky diode, has not been observed in earlier $I-V$ studies.\cite{T.Susaki, Y.Hikita} The lowering of the SBH from 1.1 eV and the deviation of n from 1 for the high doped Au/Nb:STO Schottky diodes can be attributed to several factors as the dominance of thermally assisted tunneling to electronic transport, image force lowering, generation/depletion currents within the depletion region and conduction through interface states.\cite{M.Sze}
Band bending in heavily doped Nb:STO results in narrowing of the depletion width thus promoting tunneling, accompanied by significant leakage at the M/S interface and ideality factor deviating from 1. Depletion width at zero bias, depends on the doping concentration as $W_{dep}= \sqrt {(2\epsilon_s\left(\frac{V_{i_n}}{qN_D}\right)}$ where $\epsilon_s$, $N_D$, V$_{in}$ are  semiconductor permittivity, carrier concentration and built-in potential respectively. Using $N_D$ from Fig. 2 and V$_{in}$ values from [12] and [14], we find W${dep}$ to be $\sim$ 70 nm and $\sim$ 13 nm for 0.01 wt $\%$ and 0.1 wt $\%$ Nb:STO respectively. This supports the dominance of transport processes other than thermionic emission for Au on 0.1 wt$\%$ Nb:STO Schottky diode.\cite{T.Shimizu, C.Park, D.S.Shang, A.Ruotolo}

In conclusion, electronic transport in Nb:STO single crystals for two doping densities establishes the absence of carrier freeze out upto 100 K. Electrical transport across the interface between well-characterized Nb:STO and Au were also studied. Analysis of macroscopic I-V characteristics reveal thermionic emission dominated transport for the low doped substrate whereas it deviates from such behavior for the high doped substrate. Such Schottky diodes are the building blocks of electronic devices essential for oxide electronics.

We thank S. Parui and B. J. van Wees for scientific discussions and B. Noheda and T. T. M. Palstra for use of their facilities. Technical support from J. Baas and J. G. Holstein is thankfully acknowledged. This work is supported by the Netherlands Organization for Scientific Research NWO-VIDI program.

\clearpage


\vspace*{5mm}

\begin{figure} [htb]
\includegraphics[scale=0.55]{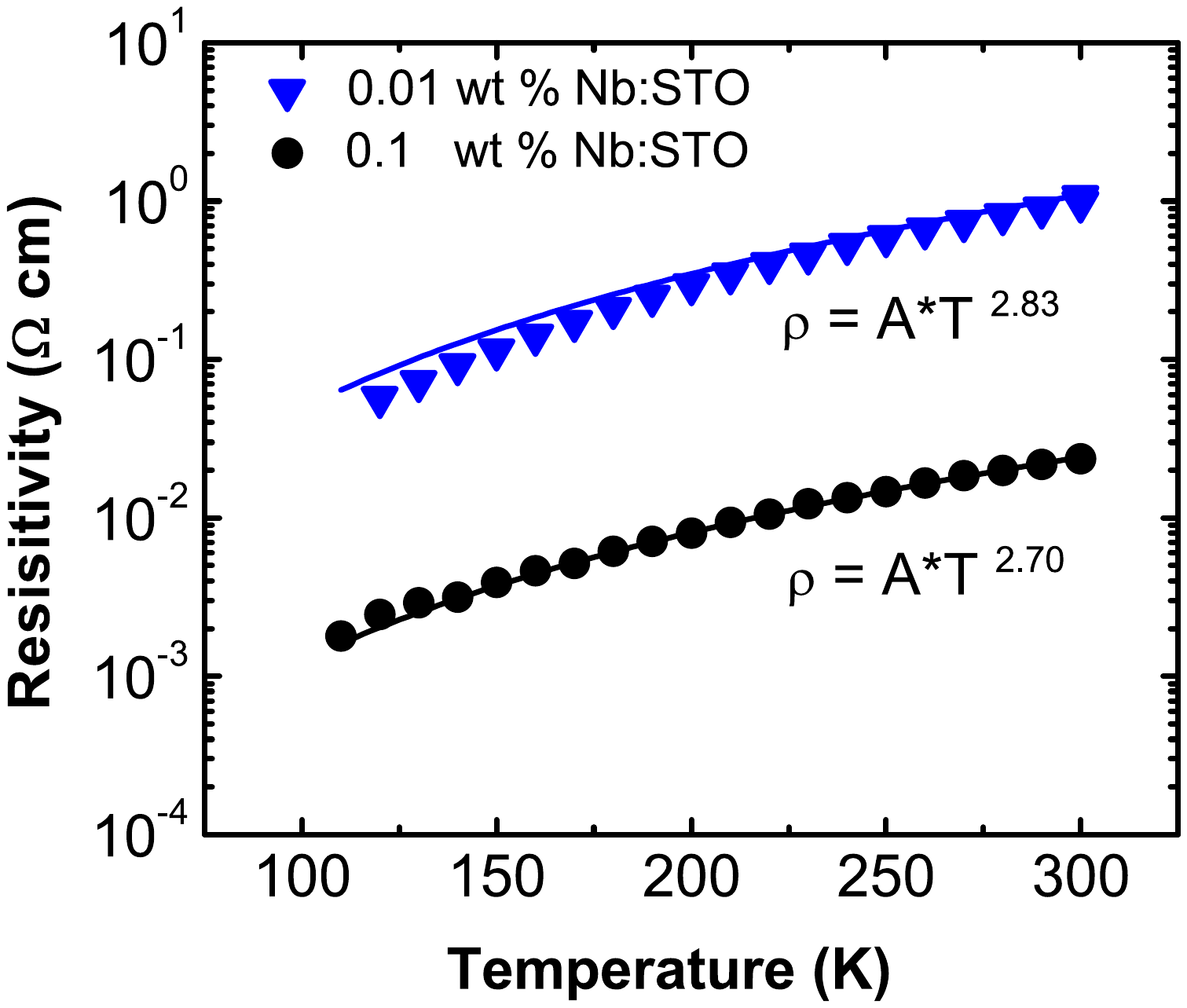}
\caption{\label{1} Temperature dependence of resistivity in Nb:STO for 0.01 and 0.1 wt $\%$ Nb doping for the temperature range 100 K to 300 K.
The solid lines represent a generic power law fitting to the experimental data.}
\end{figure}

\vspace*{5mm}

\begin{figure} [htb]
\includegraphics[scale=0.55]{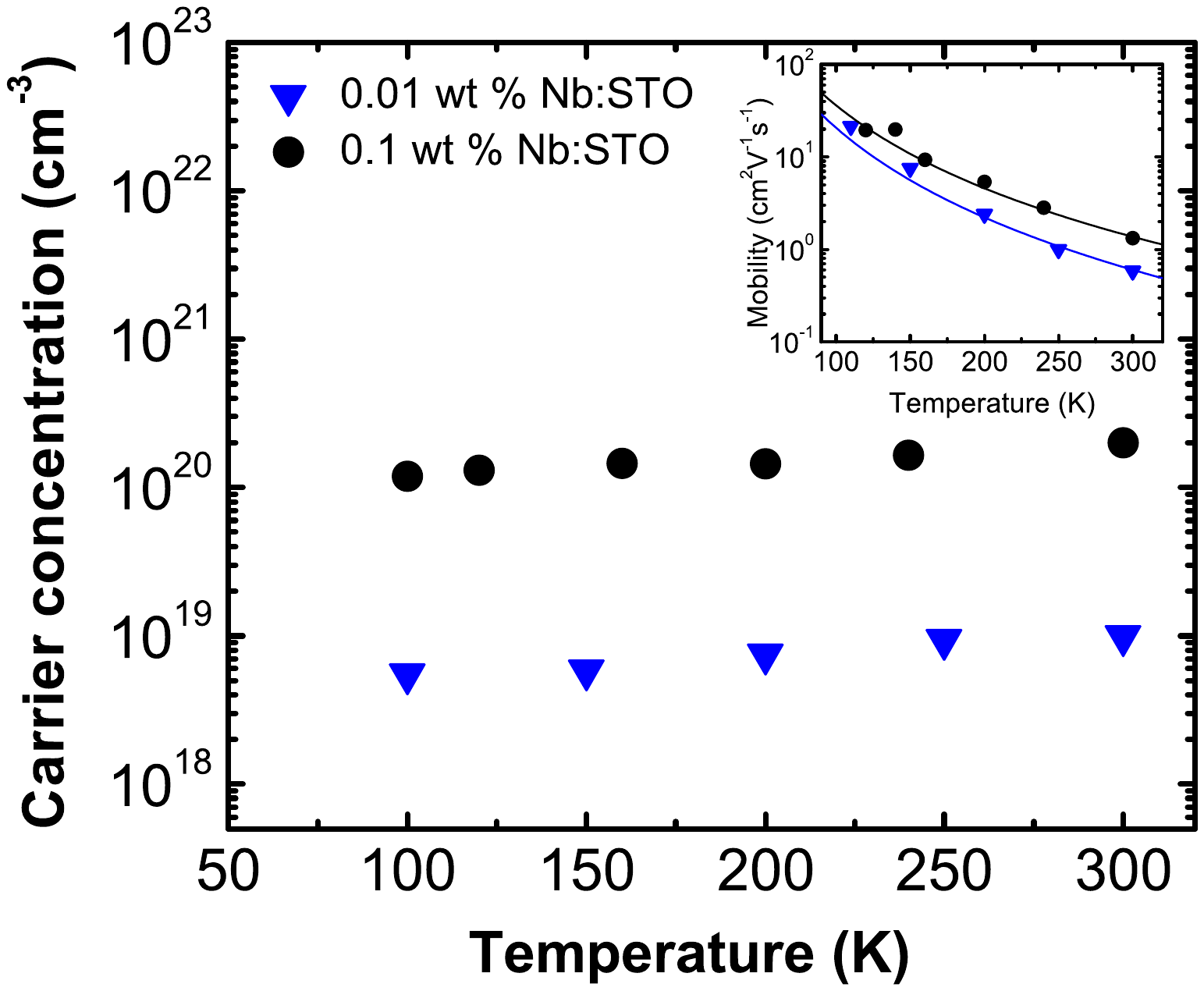}
\caption{\label{2}Temperature dependence of carrier concentration in Nb:STO for 0.01 and 0.1 wt $\%$ Nb doping for the temperature range 100 K to 300 K. Inset shows the temperature dependence of mobility, calculated from obtained resistivity and carrier concentration. The solid lines represent power law fits.}
\end{figure}

\vspace*{5mm}
                                                                                                                                      
\begin{figure} [htb]
\includegraphics[scale=0.55]{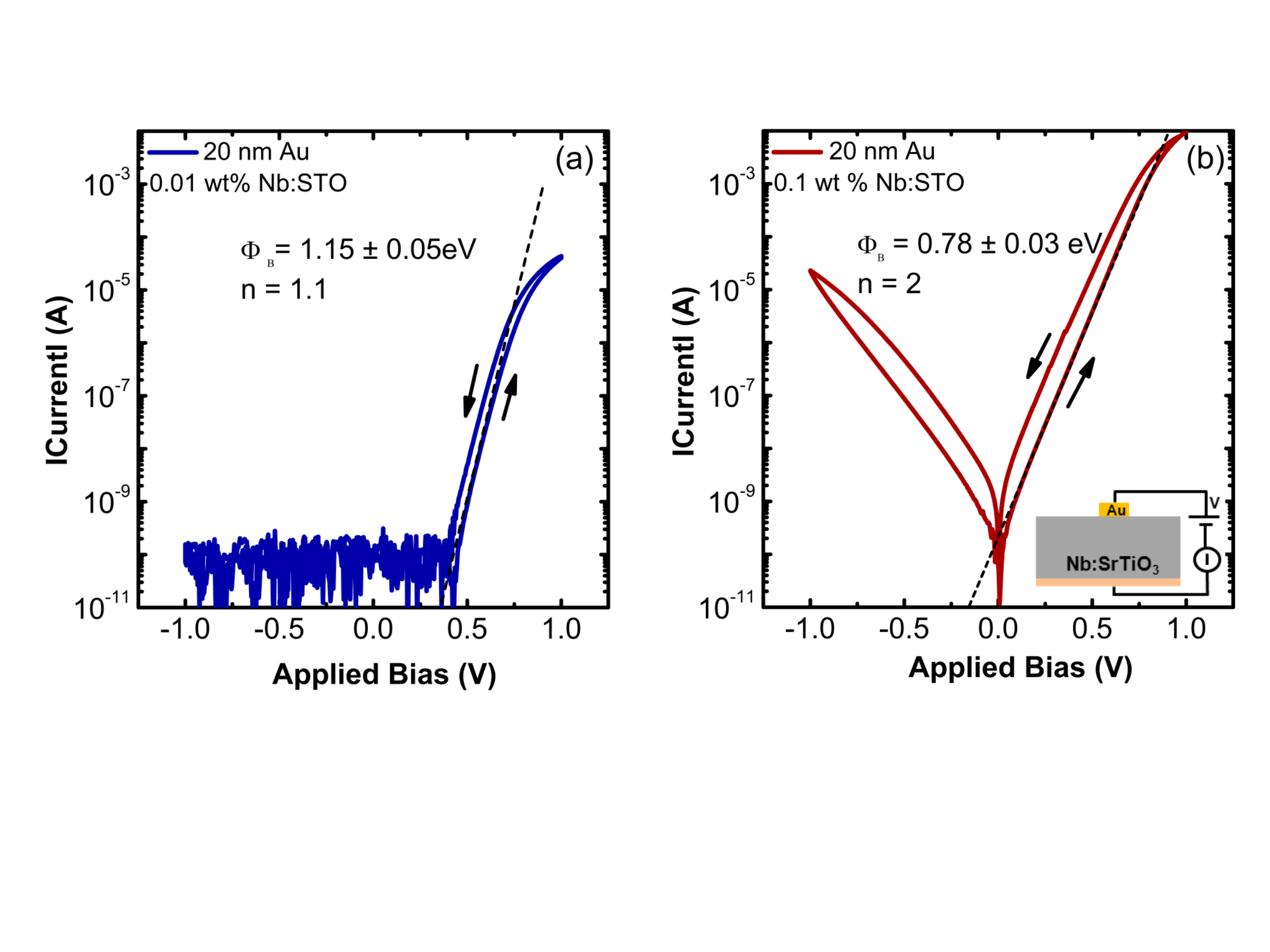}
\caption{\label{3} Current (I) - voltage (V) characteristics for Au/Nb:STO junction for (a) 0.1 and (b) 0.01 wt $\%$ Nb:STO. The up and down arrows indicate the direction of voltage sweep which yields different values of the SBH as depicted in the error bar. The dotted line represents the thermionic equation fit to the forward characteristics. Inset in (b) shows the schematic layout of the device.}
\end{figure}

\end{document}